\documentclass[superscriptaddress,twocolumn,prl,aps,showpacs]{revtex4}

\usepackage[T1]{fontenc}\usepackage[latin1]{inputenc}
\usepackage{dcolumn,graphicx,color,booktabs,microtype}
\usepackage[charter]{mathdesign}

\begin{document}
\title{Incommensurate magnetic order and dynamics induced by spinless impurities in $\bf YBa_2Cu_3O_{6.6}$}

\author{A.~Suchaneck}
\affiliation{Max Planck Institute for Solid State Research, D-70569 Stuttgart, Germany}

\author{V.~Hinkov}
\affiliation{Max Planck Institute for Solid State Research, D-70569 Stuttgart, Germany}

\author{D.~Haug}
\affiliation{Max Planck Institute for Solid State Research, D-70569 Stuttgart, Germany}

\author{L.~Schulz}
\affiliation{Physics Department and Fribourg Center for Nanomaterials, Fribourg University, CH-1700 Fribourg, Switzerland}

\author{C.~Bernhard}
\affiliation{Physics Department and Fribourg Center for Nanomaterials, Fribourg University, CH-1700 Fribourg, Switzerland}

\author{A.~Ivanov}
\affiliation{Institut Laue-Langevin, 156X, F-38042 Grenoble Cedex 9, France}

\author{K.~Hradil}
\affiliation{Institute for Physical Chemistry, University of G\"ottingen, D-37077 G\"ottingen, Germany}

\author{C.T.~Lin}
\affiliation{Max Planck Institute for Solid State Research, D-70569 Stuttgart, Germany}

\author{P.~Bourges}
\affiliation{Laboratoire L\'eon Brillouin, CEA-CNRS, CE-Saclay, F-91191 Gif-s\^ur-Yvette, France}

\author{B.~Keimer}
\email{b.keimer@fkf.mpg.de} \affiliation{Max Planck Institute for Solid State Research, D-70569 Stuttgart, Germany}

\author{Y.~Sidis}
\email{yvan.sidis@cea.fr} \affiliation{Laboratoire L\'eon Brillouin, CEA-CNRS, CE-Saclay, F-91191 Gif-s\^ur-Yvette,
France}


\begin{abstract}
We report an inelastic-neutron-scattering and muon-spin-relaxation
study of the effect of 2\% spinless (Zn) impurities on the magnetic
order and dynamics of $\rm YBa_2Cu_3O_{6.6}$, an underdoped
high-temperature superconductor that exhibits a prominent
spin-pseudogap in its normal state. Zn substitution induces static
magnetic order at low temperatures and triggers a large-scale
spectral-weight redistribution from the magnetic resonant mode at 38
meV into uniaxial, incommensurate spin excitations with energies
well below the spin-pseudogap. These observations indicate a
competition between incommensurate magnetic order and
superconductivity close to a quantum critical point. Comparison to
prior data on La$_{2-x}$Sr$_x$CuO$_{4}$ suggests that this behavior
is universal for the layered copper oxides and analogous to
impurity-induced magnetic order in one-dimensional quantum magnets.
\end{abstract}

\maketitle

Spinless impurities have provided key insights into the spin
correlations of low-dimensional quantum magnets. Research on
quasi-one-dimensional (1D) quantum-disordered magnets with a spin
gap has shown that every magnetic vacancy contributes to a Curie
term in the uniform susceptibility and generates a cloud of
antiferromagnetically correlated spins whose spatial extent is
inversely proportional to the gap \cite{Alloul09}. The striking
discovery that a dilute concentration of spinless impurities induces
antiferromagnetic long-range order in a spin-Peierls system
\cite{Hase93} can be attributed to constructive interference between
spin-polarization clouds centered at different impurities. Related
``order from disorder'' phenomena were later observed in a wide
class of quasi-1D systems and shown to exhibit universal
characteristics \cite{Azuma97,Bobroff09}. In contrast to the
comprehensive understanding that has been developed for 1D magnets,
theoretical work on spinless impurities in 2D quantum magnets has
thus far been restricted to analytical calculations at quantum
critical points \cite{Sachdev99,Hoglund07,Metlitski07,Vojta09} and
numerical simulations for specific 2D spin Hamiltonians with
quantum-disordered ground states \cite{Wessel01}. Both approaches
predict impurity-induced spin textures and order-from-disorder
phenomena not unlike those observed in 1D magnets. Experimental
tests of these predictions have, however, proven difficult because
of the paucity of physical realizations of 2D spin liquids and the
limited chemical solubility of nonmagnetic atoms in specific
compounds.

Layered copper oxides have been a fruitful testing ground for
theories of 2D quantum magnetism and for the behavior of spinless
impurities in two dimensions, despite additional complications
introduced by the presence of itinerant charge carriers and $d$-wave
superconductivity \cite{Alloul09,Hirschfeld09}. Spinless impurities
can be readily introduced into the CuO$_2$ planes by replacing
spin-1/2 copper ions by nonmagnetic zinc or lithium. A large body of
nuclear magnetic resonance (NMR) experiments on metallic and
superconducting $\rm YBa_2Cu_3O_{6+x}$ (YBCO$_{6+x}$) has
demonstrated that such impurities generate a Curie term in the
uniform susceptibility and a staggered spin polarization cloud on
neighboring Cu sites, in close analogy to the 1D systems discussed
above \cite{Alloul09}. NMR experiments on other families of cuprates
are much more limited, because intrinsic disorder due to dopant
atoms broadens the NMR lines, so that the effect of Zn impurities is
masked by native disorder \cite{Bobroff02}. Inelastic neutron
scattering (INS) experiments on both YBCO$_{6+x}$ and
La$_{2-x}$Sr$_x$CuO$_{4}$ (LSCO) have revealed Zn-induced low-energy spin
excitations, in qualitative agreement with the NMR results. In
contrast to the 1D systems \cite{Bobroff09}, however, a universal
picture of the impurity-induced magnetism has not yet emerged in the
2D cuprates, because the spatial character of the impurity-induced
low-energy spin dynamics in both materials appeared to be rather
different: commensurate antiferromagnetic in YBCO$_{6+x}$
\cite{Sidis96,Kakurai93} and incommensurate (IC) in LSCO
\cite{Kimura03,Kofu05}. Moreover, Zn-induced static incommensurate
magnetism, which was observed in LSCO
\cite{Kimura03,Kofu05,Panagopoulos04} and interpreted as evidence of
pinning of fluctuating stripes \cite{Kivelson03}, has thus far not
been detected in YBCO$_{6+x}$. For this reason, and because LSCO exhibits
static IC magnetic order at some doping levels even in the absence
of spinless impurities \cite{Kimura99,Kofu09}, the universality of
this ``order from disorder'' phenomenon has been questionable. These
considerations have motivated us to revisit the effect of spinless
impurities on the magnetic correlations of YBCO$_{6.6}$, a material
with minimal native disorder that is well known for its large
normal-state spin-pseudogap \cite{Alloul09}.

The experiments were performed on an array of detwinned, co-aligned
$\rm YBa_2(Cu_{0.98}Zn_{0.02})_3O_{6.6}$ (henceforth
YBCO$_{6.6}$:Zn) single crystals of total mass 650 mg and
superconducting $T_\mathrm{c} =30$ K. The Zn content was determined
by energy-dispersive x-ray analysis and inductively-coupled plasma
spectroscopy. A twin domain population ratio of 4:1 enabled us to
discriminate spin excitations propagating along the $a$- and
$b$-axes in the CuO$_2$ planes, which is important in view of the
recent observation of uniaxial, incommensurate magnetism in
YBCO$_{6.45}$ \cite{Hinkov08,Haug09}. The INS measurements were
carried out on the spectrometers IN8 (Institut Laue-Langevin,
Grenoble), 4F2, 1T , 2T (Orph\'ee, Laboratoire L\'eon Brillouin,
Saclay), and PUMA (FRM-II, Garching), using either a
pyrolytic-graphite (PG) (002) or a Si (111) monochomator in
combination with a PG (002) analyzer. PG filters extinguished
higher-order contaminations of the beam. The final wave vector was
fixed to 4.1 \AA$^{-1}$ for excitation energies $E \geq 25$ meV, and
to 2.662 \AA$^{-1}$ or 1.97 \AA$^{-1}$ for $E < 25$ meV. Reference
measurements were performed on a Zn-free YBCO$_{6.6}$ array with
$T_\mathrm{c} =61$ K described previously \cite{Hinkov07}. Data on
both arrays were compared by calibration measurements on the optical
phonon at 42.5 meV following previous work \cite{Fong00}, and by
performing measurements under identical conditions. The wave vector
transfers ${\bf Q}=(H,K,L)$ are given in units of the
reciprocal-lattice vectors ${\bf a}^{\star}$, ${\bf b}^{\star}$, and
${\bf c}^{\star}$. ${\it Q}=(H,K)$ stands for 2D projections of {\bf
Q}, and $\rm {\it Q}_{\it AF} =(0.5,0.5)$ is the wave vector
characterizing antiferromagnetism in undoped YBCO$_{6.0}$.

\begin{figure}[t]
\includegraphics[width=0.9\columnwidth,angle=0]{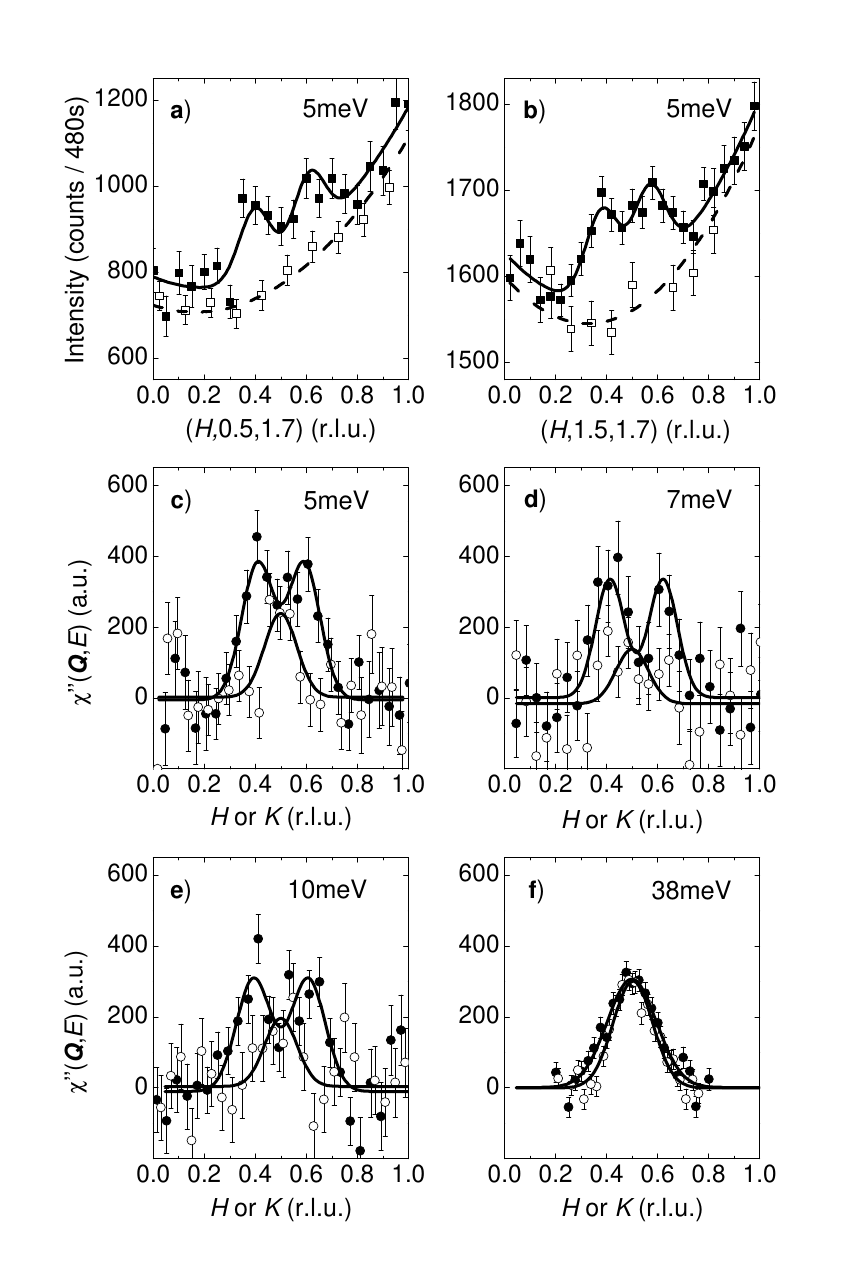}
\caption {Constant-energy scans in YBCO$_{6.6}$:Zn
at $T = 10$ K. (a,b) Uncorrected scans
along (a) ${\bf Q}=(H, 0.5, 1.7)$ and (b) $(H, 1.5, 1.7)$. The extraneous background was determined by rotating the sample by a) 30$^{\circ}$, b) 10$^{\circ}$ and repeating the scans (open symbols). (c-f) Background-corrected scans along ${\bf Q}=(H, 1.5, 1.7)$ (full symbols) and $(1.5, K, 1.7)$ (open symbols) at $E = 5$, 7, 10, and 38 meV.}
\end{figure}

Figure 1a,b shows INS scans taken along the $a$-axis of
YBCO$_{6.6}$:Zn at $E=5$ meV, well below the spin pseudogap of
pristine YBCO$_{6.6}$. A pair of well separated IC peaks
symmetrically displaced from ${\it Q}_{AF}$ is observed consistently
in two Brillouin zones on top of an extraneous background arising
from the sample mount. The background is independent of the sample
orientation and can be readily subtracted (Fig. 1c). The signal is
absent at $L = 0$ and vanishingly small at large $|{\bf Q}|$, and
can thus be ascribed to acoustic spin fluctuations \cite{Fong00}.
Very similar peaks are also observed at $E=7$ and 10 meV (Fig.
1d,e). As scans along the $b$-axis are peaked at ${\it Q}_{\it AF}$,
the excitations are characteristic of uniaxial IC magnetism. Since
low-energy spin excitations of comparable intensity are not observed
in control experiments on pristine YBCO$_{6.6}$ (see below), they can
be identified as a result of Zn substitution.

The incommensurability of the impurity-induced spin excitations
eluded detection in previous INS experiments on YBCO$_{6+x}$:Zn
\cite{Sidis96,Kakurai93} for three reasons: First, the signal from
the previously investigated twinned crystals consists of equal
contributions from two twin domains, so that the IC character is
less apparent. Second, earlier studies were performed in the
scattering plane spanned by the vectors ${\bf Q} = (1, 1, 0)$ and
$(0, 0, 1)$, which does not include the IC wave vector. Third, the
effective momentum-space resolution in the latter configuration is
poorer than in the present experiment, because the long, vertical
axis of the resolution ellipsoid is parallel to the $\rm
CuO_2$-planes.

The INS profiles were fitted to Gaussians centered at ${\it Q}_{\it
AF} \pm (\delta, 0)$ (lines in Figs. 1c-e). In the range $E= 5- 10$
meV, both the incommensurability $\delta = 0.10 \pm 0.05$ and the
intrinsic half-width-at-half-maximum $\Delta Q = 0.055 \pm 0.025$
extracted from these fits were found to be energy independent within
the error. The incommensurability is consistent with the one that
characterizes excitations above the spin-pseudogap of pristine
YBCO$_{6.6}$ \cite{Hinkov07}, but significantly larger than the one
observed in YBCO$_{6.45}$ \cite{Hinkov08}. This confirms that the Zn
content and the doping level are independent control parameters
\cite{Alloul09}. Models according to which Zn impurities
reduce the hole doping level
\cite{Kaplan02,Abrikosov03} are inconsistent with our data. The
intrinsic momentum width translates into a length scale of $\xi =
2.9 a$, where $a$ is the in-plane lattice spacing. This represents a
lower bound on the intrinsic coherence length of the spin
fluctuations, because inhomogeneous broadening (arising from the
doping dependence of $\delta$ in combination with a possible slight
nonuniformity of the doping level) may also contribute to $\Delta
Q$. In any case, the value of $\xi$ determined in this way is in
good agreement with NMR data \cite{Ouazi04} and comparable to $5.8
a$, the mean distance between Zn impurities in the CuO$_2$ planes.

\begin{figure}[t]
\includegraphics[width=7.0cm,angle=0]{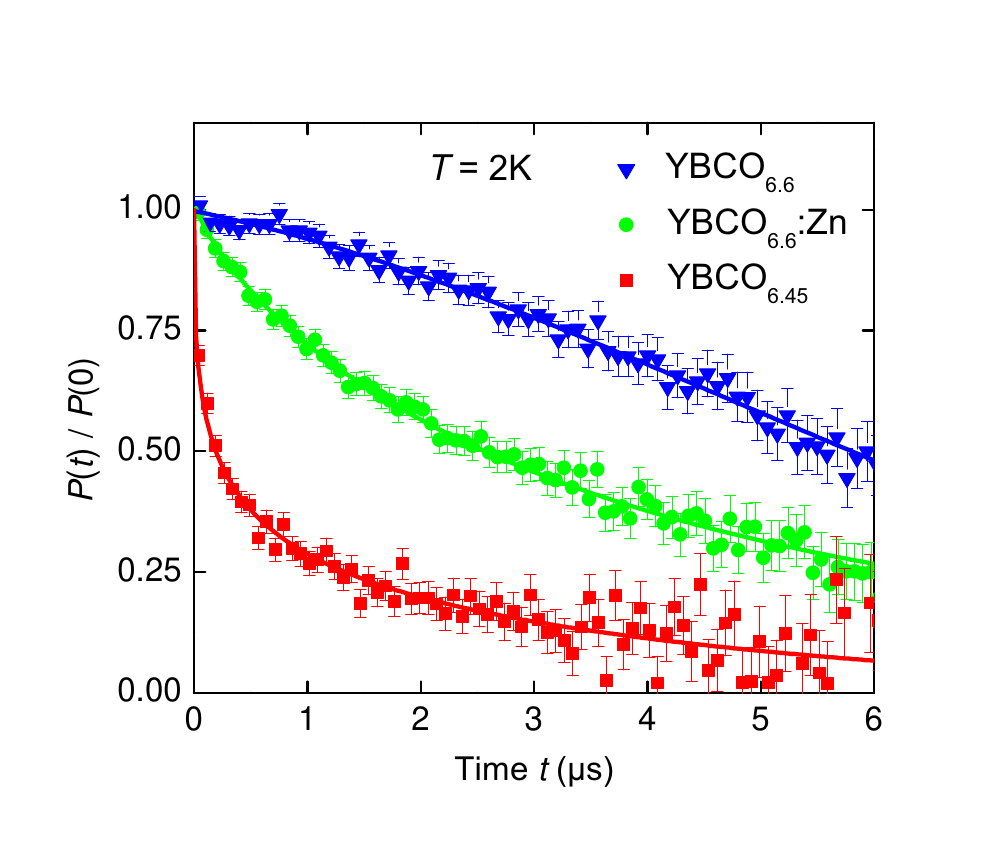}
\caption { (color online) Zero-field muon-spin relaxation data for YBCO$_{6.6}$:Zn
and reference samples $\rm YBa_2Cu_3O_{6.6}$ and $\rm YBa_2Cu_3O_{6.45}$ at $T = 2$ K. The lines are the results of fits
described in the text.}
\end{figure}

In view of the sharp excitation peaks, we have investigated the
presence of magnetic order in YBCO$_{6.6}$:Zn. Since a quasielastic
neutron signal like the one previously observed in YBCO$_{6.35}$
(Ref. \onlinecite{Stock06}) and YBCO$_{6.45}$ (Ref.
\onlinecite{Haug09}) could not be detected above background
\cite{note}, we performed muon spin relaxation ($\mu$SR) measurements,
which are well suited to search for weak magnetic order. The
experiments were performed with the GPS setup at the $\pi$m3
beamline at the Swiss muon source S$\mu$S at the Paul Scherrer Institute,
Villigen, Switzerland. The time-resolved $\mu$SR spectra of
YBCO$_{6.6}$:Zn as well as reference spectra of Zn-free YBCO$_{6.6}$
and YBCO$_{6.45}$ (Fig. 2) were fitted to the relaxation function
$P(t)=P(0) \exp[(- \lambda t)^\beta ] \times KT$ that consists of
the product of a stretched exponential and a Kubo-Toyabe ($KT$)
function, which account for the contribution of the electronic and
nuclear magnetic moments, respectively. For YBCO$_{6.6}$ the
relaxation is dominated by the nuclear moments, suggesting the
absence of static or slowly fluctuating electronic moments. The
YBCO$_{6.6}$:Zn spectra exhibit a more rapid
depolarization at temperatures below $\sim 10$ K (Fig. 2), which is
a signature of dominant contributions from electronic moments.  The
absence of a precession frequency indicates that these electronic
moments are either static and disordered or slowly fluctuating. The
obtained values of $\beta \sim 0.4$ for YBCO$_{6.45}$ and $\beta
\sim 0.7$ for YBCO$_{6.6}$:Zn are characteristic of a spatially
nonuniform distribution of the magnetic moment amplitude or
fluctuation rate \cite{Panagopoulos04}. These observations confirm
prior $\mu$SR results \cite{Alloul09,Mendels94,Bernhard97} according
to which spinless impurities extend the range of magnetic order in
the phase diagram of YBCO$_{6+x}$ up to $x \sim 0.7$. The low-energy
INS data (Fig. 1) and the close similarity between the $\mu$SR
spectra of YBCO$_{6.6}$:Zn and those of pure YBCO$_{6.45}$ (Fig. 2)
now indicate that this order is incommensurate, in contrast to the
commensurate antiferromagnetism suspected on the basis of prior INS
data on YBCO$_{6+x}$:Zn \cite{Sidis96,Kakurai93}.

\begin{figure}[t]
 \includegraphics[width=0.9\columnwidth,angle=0]{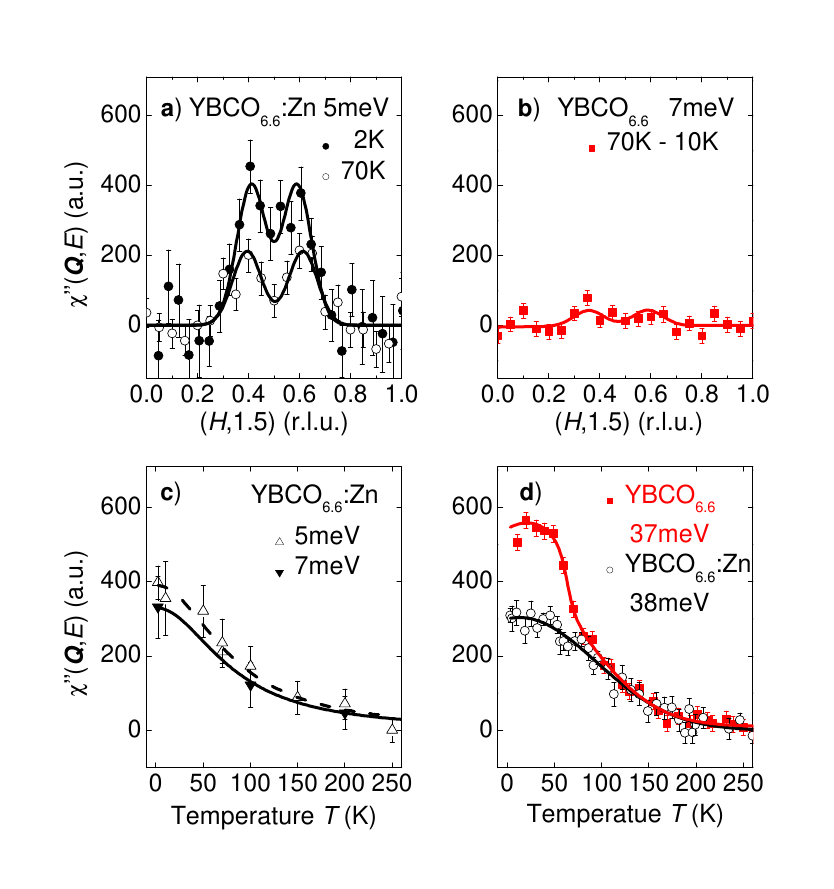}
\caption { (color online) Temperature dependence of the spin fluctuation intensity. (a,b) Scans of the dynamical
magnetic susceptibility, $\chi '' ({\bf Q}, E)$ (derived from the INS data by correcting for the detailed-balance
factor), around ${\bf Q} = (0.5, 0.5, 1.7)$ in Zn-substituted and Zn-free YBCO$_{6.6}$ at energies and
temperatures marked in the legend. The data were background-corrected as described in Fig. 1. The lines are the results
of fits described in the text. (c) Amplitude of $\chi ''$ in YBCO$_{6.6}$:Zn
at $E=5$ and 7 meV extracted from fits to constant-energy scans. (d) Amplitude of $\chi''$ for YBCO$_{6.6}$:Zn (YBCO$_{6.6}$)
at $E = 38$ (37) meV. The lines are
guides-to-the-eye. }
\end{figure}

We now describe the behavior of the spin system at higher temperatures and excitation energies. At $T=70$ K (that is,
above $T_\mathrm{c}$ of both YBCO$_{6.6}$ and YBCO$_{6.6}$:Zn), low-energy IC fluctuations persist in YBCO$_{6.6}$:Zn
without significant thermal broadening (Fig. 3a), and remain much more intense than those in YBCO$_{6.6}$ (Fig. 3b).
The onset temperature of the IC intensity in YBCO$_{6.6}$:Zn is $\sim
150$ K (Fig. 3c), comparable to the onset of the spin-pseudogap in YBCO$_{6.6}$ previously documented by NMR
\cite{Alloul09}. This shows that Zn substitution restores low-energy spin fluctuations not only in the superconducting
state, but also in the pseudogap state.
The characteristic temperature for the Zn-induced enhancement of the spin
dynamics is also comparable to the onset temperature of IC spin fluctuations in YBCO$_{6.45}$, which provides evidence
for an electronic nematic state \cite{Hinkov08}.

A converse spectral-weight shift is observed at higher energies. Pure YBCO$_{6.6}$ exhibits a sharp resonant mode with
energy $E_{res} = 38$ meV in the superconducting state \cite{Fong00,Hinkov07}, which is manifested by a sharp
enhancement of the spin fluctuation intensity for $E \sim E_{res}$ below $T_\mathrm{c}$ (Fig. 3d). While
YBCO$_{6.6}$:Zn exhibits a similar narrowing of the magnetic response around $Q_{AF}$ and $E = 38$ meV as pure
YBCO$_{6.6}$ (Fig. 1f), followed by a dispersion away from $Q_{AF}$ at higher energies (not shown), the
superconductivity-induced anomaly is obliterated by Zn substitution (Fig. 3d). This is consistent with previous INS
studies close to optimal doping that had demonstrated progressive weakening of the resonant mode with increasing Zn
content without significant renormalization of its characteristic energy \cite{Fong99,Sidis00,Sidis01}.

\begin{figure}[t]
 \includegraphics[width=7.0cm,angle=0]{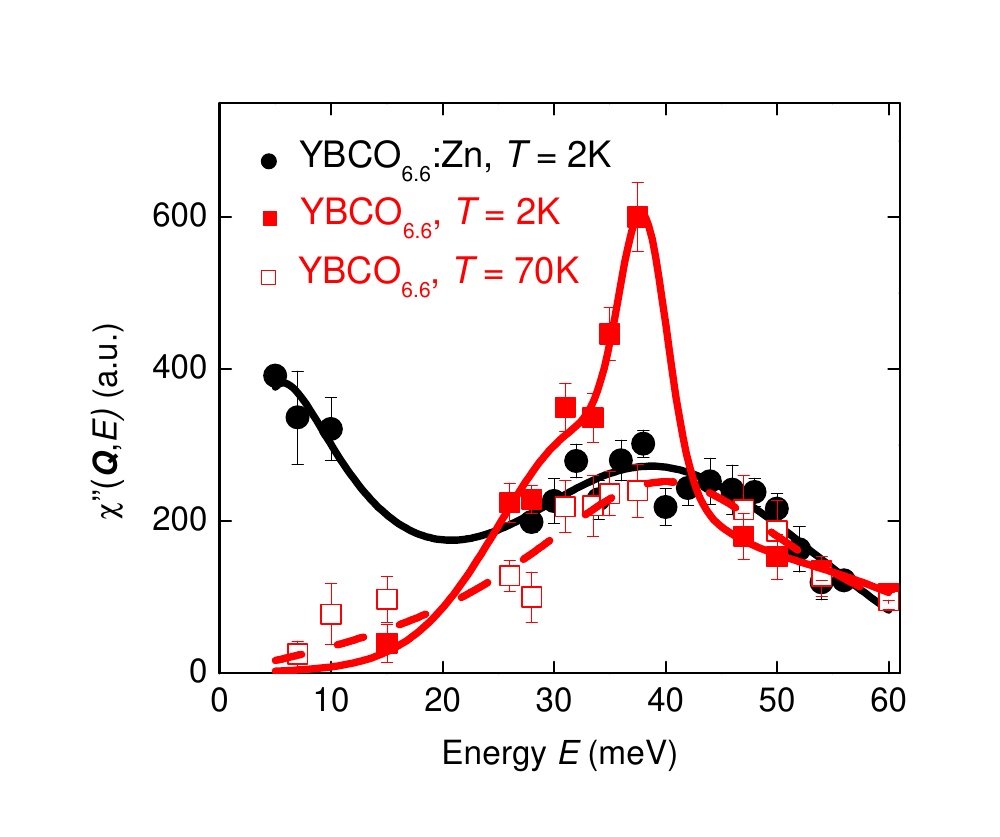}
\caption { (color online)  Energy dependence of the peak susceptibility $\chi ''$ at $T=2$ K extracted from
constant-energy scans on YBCO$_{6.6}$:Zn and YBCO$_{6.6}$. Normal-state data on pure YBCO$_{6.6}$ are shown for comparison. The lines are guides-to-the-eye. }
\end{figure}

Figure 4 provides a synopsis of the spin excitation spectra of
YBCO$_{6.6}$ and YBCO$_{6.6}$:Zn. The YBCO$_{6.6}$:Zn spectrum at
low temperatures is closely similar to the normal-state spectrum of
YBCO$_{6.6}$, except for the strongly enhanced intensity at low
energies that goes along with the appearance of IC magnetic order.
The low-temperature spectrum of YBCO$_{6.6}$, on the other hand, is
dominated by the sharp magnetic resonant mode at 38 meV, which is a
signature of the superconducting state \cite{Fong00}. Fingerprints
of both forms of collective electronic order are thus clearly
apparent in the spin excitation spectrum, and the spectral-weight
redistribution induced by Zn substitution demonstrates that spinless
impurities nucleate static IC magnetism at the expense of
superconductivity. This conclusion is consistent with the behavior
expected in the vicinity of a quantum critical point separating both
phases, and with prior $\mu$SR studies that had indicated a local
suppression of the superfluid density around Zn impurities
\cite{Bernhard96,Nachumi96}. The two-phase competition also explains
the extreme sensitivity of the magnetic resonant mode to a minute
concentration of Zn impurities \cite{Fong99,Sidis00,Sidis01}, which
is difficult to explain in the framework of the random-phase
approximation commonly used to describe the origin of this mode
\cite{Bulut00}.

In conclusion, our experiments on the prototypical spin-pseudogap
material YBCO$_{6.6}$ have shown that spinless impurities induce
incommensurate magnetic order and low-energy spin excitations with a
correlation range comparable to the mean distance between the
impurities. In conjunction with prior data on LSCO
\cite{Kimura03,Kofu05}, this indicates a universal
order-from-disorder phenomenology for copper oxides with 2D
electronic structure, closely analogous to the one previously
established for quasi-1D quantum magnets
\cite{Hase93,Azuma97,Bobroff09}. In contrast to the latter systems,
however, the spin excitations in the layered copper oxides are
incommensurate. Our demonstration that the incommensurability in the
YBCO$_{6+x}$ system is independent of impurity concentration and controlled
by the density of itinerant charge carriers, as previously shown for
LSCO \cite{Kimura03,Kofu05,Kimura99,Kofu09}, explains the difference
to the Mott-insulating 1D systems and further underscores the
universality of impurity-induced spin correlations in
low-dimensional quantum magnets.

We acknowledge discussions with E. Fradkin, P.J. Hirschfeld, H.Y. Kee, S.A. Kivelson, and O.P. Sushkov, and
financial support by the DFG under grant FOR538.



\begin{thebibliography}{99}
\bibitem{note} Note that the intensity of the quasielastic neutron signal reported for YBCO$_{6.45}$ \cite{Hinkov08,Haug09} is at the edge of the detection limit.
Since the volume of the present YBCO$_{6.6}$:Zn sample is factor of
three smaller, and the ordered moment is weaker (Fig. 3), the
quasielastic signal is expected to be below the detection limit, in
agreement with the data.
\end{thebibliography}
\end{document}